\documentclass[]{spie}  

\usepackage{amsmath,amsfonts,amssymb}
\usepackage{graphicx}
\usepackage[colorlinks=true, allcolors=blue]{hyperref}

\title{ESOFinder: an LLM-powered tool to help users navigate ESO documentation}

\author[a]{Paula S\'anchez-S\'aez}
\author[a]{Claudio Reinero}
\author[a]{Miguel Vioque}
\author[a]{Markus Wittkowski}
\author[a]{Marina Rejkuba}
\author[a]{Martino Romaniello}
\author[a]{Ashley Barnes}
\author[a]{John Pritchard}
\author[a]{Michaël Marsset}

\affil[a]{European Southern Observatory, Karl-Schwarzschild-Strasse 2, 85748 Garching bei München, Germany}

\authorinfo{Further author information: (Send correspondence to P.S.S.)\\P.S.S.: E-mail: paula.sanchezsaez@eso.org}

\begin{document} 
\maketitle

\begin{abstract}
The large amount and diversity of documentation available for users of the European Southern Observatory (ESO) -- spanning the full observing lifecycle from proposal preparation and observation planning to data reduction and archival access -- makes it increasingly challenging for the astronomical community to efficiently find relevant information. To address this, we have developed ESOFinder, an in-house chatbot powered by Large Language Models (LLMs) and Retrieval-Augmented Generation (RAG). ESOFinder integrates public information from instrument manuals, phase 1/2/3 documentation, data reduction pipeline manuals, the ESO Knowledge Base, and key web resources (spanning more than 3500 links and over 100 manuals) to provide concise, context-aware, and reference-linked answers to user queries about proposal/observation preparation, data retrieval, and data processing. Built on open-source LLMs (e.g., Mistral AI models) running on a local server, ESOFinder ensures data privacy, transparency, and complete control over the knowledge base. Its multi-step architecture allows verification of retrieved documents and generated answers, reducing the risk of hallucinations and improving the reliability of responses compared to commercial tools. The current version of ESOFinder is being tested internally at ESO to evaluate its performance, assess its integration with internal workflows, and identify limitations in coverage and accuracy. These tests will guide further improvements, including the incorporation of additional documentation, and enhanced retrieval strategies. Ultimately, ESOFinder aims to become an interface for users to navigate ESO’s complex documentation ecosystem and to support both staff and community astronomers in their daily tasks.
\end{abstract}


\section{INTRODUCTION}
\label{sec:intro}  

Modern astronomical observatories generate vast and continuously evolving documentation ecosystems that users must navigate to successfully plan and execute their observing programmes. At the European Southern Observatory (ESO), the documentation of La Silla Paranal Observatory spans more than 3500 links across ESO science webpages\footnote{\url{https://www.eso.org/sci.html}} and Knowledge base\footnote{\url{https://support.eso.org/en-GB/kb}}, and over 100 manuals distributed as PDF documents, totalling over 6.2 million words (roughly eight times the length of the King James Bible, or the equivalent of over 17 days of continuous reading at average pace). It encompasses a broad and heterogeneous set of resources covering the full lifecycle of an observational programme: from Phase 1, the proposal submission stage in which astronomers define their science case, describe the programme feasibility, and provide a target list, instrument configuration, and time request for evaluation and allocation; through Phase 2, the observation preparation and execution stage in which accepted proposals are translated into detailed observing blocks (OBs) specifying telescope pointings, instrument setups, and execution constraints, and in which the observing programmes are executed and monitored; to Phase 3, the data product submission stage in which principal investigators contribute science-ready reduced data back to the ESO archive for public release. Beyond these phase-specific resources, users must also navigate instrument user manuals, data reduction pipeline documentation, and archive interface instructions. Locating precise, up-to-date information within this extensive corpus represents a significant and growing challenge, both for the astronomical community and for ESO staff, who assist users on a daily basis and must maintain curated, current documentation across all of these resources. Motivated by this need, we present ESOFinder, a question-answering system designed to assist ESO users by leveraging recent advances in natural language processing (NLP), specifically large language models (LLMs) and retrieval-augmented generation (RAG).

An LLM is an artificial intelligence (AI) model designed to understand and generate human-like text. Most modern LLMs are based on the Transformer architecture introduced by Vaswani et al. (2017) \cite{Vaswani17}, and are trained on large, diverse text corpora that enable them to perform tasks such as translation, summarisation, and open-ended question answering. The term ``large" refers to the substantial number of parameters these models contain (typically in the billions), which allows them to capture complex linguistic patterns and generate coherent responses. While general-purpose LLMs such as those powering ChatGPT\cite{OpenAI23}, Gemini\cite{geminiteam2023gemini}, or Claude\cite{anthropic2024claude3} have demonstrated impressive capabilities, they present several limitations: their training data has a fixed cutoff date and cannot incorporate domain-specific or frequently updated knowledge; they are prone to generating hallucinated answers, as they are designed to always produce a response rather than acknowledge uncertainty; and their use may raise privacy and security concerns when sensitive or confidential information is involved. These limitations are particularly pronounced in an observatory context, where documentation is continuously revised to reflect new instrument capabilities, operational policies, and observing modes. Moreover, a substantial amount of outdated information persists on the Internet (like legacy instrument manuals or webpages from previous observing periods), making it difficult for general-purpose LLMs to distinguish current guidelines and policies from obsolete ones. To address this, ESOFinder employs RAG\cite{Lewis20}, an approach that augments a generative model with an external retrieval mechanism. In RAG, when a user submits a query, a retrieval component first searches a curated knowledge base to identify relevant documents; this retrieved content is then provided as context to the generative model, which synthesizes a response grounded in both its pre-trained knowledge and the dynamically retrieved information. This three-stage pipeline -- retrieve, augment, generate -- makes RAG particularly well-suited to knowledge-intensive and domain-specific applications such as observatory user support, where accuracy and timeliness of information are critical.

In this paper, we introduce ESOFinder, an in-house RAG-based chatbot developed within the User Support Department (USD) at ESO Headquarters. We describe its design, including the hybrid retrieval pipeline, the answer-generation strategy, and the multi-step architecture used to reduce hallucinations and improve response reliability. We also present a quality assessment of the tool based on feedback collected from ESO staff during internal testing. Section~\ref{sec:architecture} describes the system architecture; Section~\ref{sec:evaluation} presents the evaluation results; and Section~\ref{sec:conclusions} summarises our findings and outlines future directions.

\section{ESOFinder architecture}\label{sec:architecture}

ESOFinder is composed of two operationally distinct phases: an \emph{offline ingestion pipeline} that constructs the searchable knowledge base from ESO documentation, and an \emph{online inference pipeline} that handles user queries
at runtime. The following sections provide a high-level overview of the system components and their interactions.

\subsection{Document Ingestion Pipeline}
\label{sec:ingestion}

Before the system can answer questions, all relevant documentation must be processed and indexed.  This offline pipeline transforms heterogeneous source material (both PDF manuals and web-based content) into a structured, searchable knowledge base through three main steps.
 
\textbf{Content extraction and parsing:}
ESO documentation spans a wide range of formats and topics. PDF documents are parsed into structured text that preserves the original heading hierarchy and layout. Web-based documentation is scraped and converted into a consistent plain-text representation. All content is organised into domain-specific collections covering proposal preparation, instrument configuration, data delivery and access, data reduction, and general observatory support, with dedicated coverage for La~Silla Paranal Observatory operations. ALMA documentation is maintained as a separate collection, as its instrumentation, operating modes, and documentation conventions differ substantially from those of other ESO facilities.
 
\textbf{Chunking and contextual enrichment:}
The extracted text is split into overlapping chunks using a hierarchical splitting strategy that respects document structure: sections are first divided at heading boundaries before a secondary token-level split is applied to longer segments, keeping a maximum of approximately 1,000 tokens (roughly 750 words) per chunk with a small overlap between adjacent chunks. This ensures that chunk boundaries align with natural document divisions rather than arbitrary character counts. In addition, we apply a contextual retrieval\footnote{\url{https://www.anthropic.com/engineering/contextual-retrieval}} approach, where each chunk is passed to the LLM together with a summary of its parent document, and the model is asked to generate a short description situating the chunk within its broader context. That enriched description is then added to the chunk before embedding, improving the quality of vector search for passages that would otherwise appear ambiguous when read in isolation.

\textbf{Dual indexing:}
Each enriched chunk is indexed in two complementary ways. Dense vector embeddings are produced with a sentence-transformer model (\texttt{nomic-embed-text-v1}~\cite{nussbaum2024nomic}) and stored in a ChromaDB\footnote{\url{https://github.com/chroma-core/chroma}} vector store, with separate collections per documentation domain. Simultaneously, all chunks are indexed in an Elasticsearch\footnote{\url{https://github.com/elastic/elasticsearch}} BM25~\cite{BM25} keyword index. The rationale for maintaining both representations is discussed in Section~\ref{sec:retrieval}. The ingestion step also supports zero-downtime updates: a rebuilt index is first written to a staging area and then instantly replaces the existing one, so the live service continues answering queries uninterrupted while documentation is updated.

\subsection{Hybrid Retrieval and Reranking}
\label{sec:retrieval}

\begin{figure} [ht]
\begin{center}
\begin{tabular}{c} 
\includegraphics[width=0.7\textwidth]{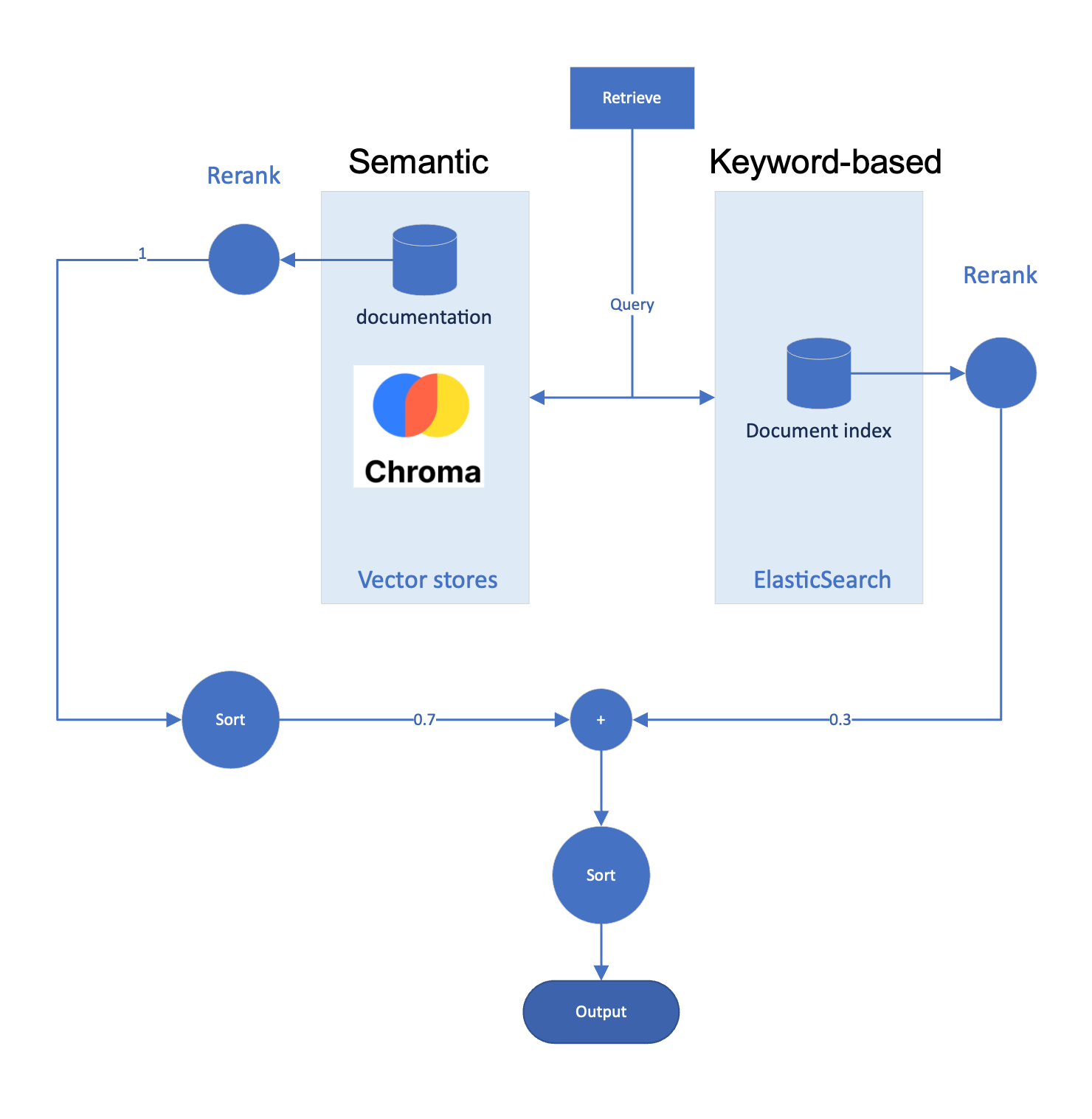}
\end{tabular}
\end{center}
\caption
{ \label{fig:rag} 
ESOFinder RAG strategy summary, including semantic and keyword-based document retrieval and reranking strategy.}
\end{figure}

When a query arrives, the system retrieves candidate documents through a three-stage pipeline designed to maximise both recall and precision. A summary of the RAG strategy is presented in Figure~\ref{fig:rag}, and further explained bellow.

\textbf{Parallel hybrid search:}
The user's question is submitted simultaneously to two search systems, shown as the \texttt{Semantic} and \texttt{Keyword-based} branches in Figure~\ref{fig:rag}.
The \texttt{Semantic} branch uses vector search: the query is converted into a dense embedding and compared against the chunk embeddings pre-computed during the documentation ingestion and stored in the ChromaDB collections (Section~\ref{sec:ingestion}). Passages are ranked by cosine similarity\cite{manning2008introduction}, a measure of how closely their meaning matches the query, allowing the system to retrieve relevant content even when the wording differs from the original question. To avoid returning near-duplicate passages, the top results are further refined using Maximal Marginal Relevance (MMR~\cite{Carbonell1998TheUseOM}), which balances relevance to the query against diversity among the selected passages.
The \texttt{Keyword-based} branch uses BM25~\cite{BM25}, a classical ranking function that scores passages according to the frequency and specificity of query terms appearing in the text. This makes it particularly effective for exact matches on technical vocabulary such as instrument names, acronyms, and OB templates, terms that semantic search may fail to distinguish because they do not separate well in embedding space. Combining both methods is a deliberate design choice: semantic search captures meaning and handles paraphrasing, while keyword search ensures that precise technical terms are not missed. Each branch independently selects 16 candidate documents, which are then merged for reranking.

\textbf{Cross-encoder reranking:}
The candidate document chunks selected in the previous step are reranked by a dedicated cross-encoder model (\texttt{BAAI/bge-reranker-base}\footnote{\url{https://huggingface.co/BAAI/bge-reranker-base}}). Unlike the embedding-based retriever, which scores documents independently, the cross-encoder evaluates each (query, passage) pair jointly, producing a more accurate relevance estimate. 
 
\textbf{Weighted score combination:}

Results from both search branches are merged into a single ranked list using a weighted scoring formula. Each passage receives a final score that combines two components: a semantic score, based on how well the passage matches the meaning of the query, and a keyword score, based on how well it matches the exact terms used. The semantic component is weighted more heavily ($\alpha = 0.7$), reflecting its generally stronger retrieval performance:

\begin{equation}
  s_{\mathrm{final}}
    = \alpha \cdot s_{\mathrm{reranker}}^{\mathrm{vec}} \cdot w_{\mathrm{coll}}
    + (1-\alpha) \cdot \frac{0.1 \cdot s_{\mathrm{reranker}}^{\mathrm{BM25}}}{r_{\mathrm{BM25}} + 1},
  \label{eq:fusion}
\end{equation}
 
\noindent where $s_{\mathrm{reranker}}^{\mathrm{vec}}$ and $s_{\mathrm{reranker}}^{\mathrm{BM25}}$ are the cross-encoder relevance scores for candidates retrieved by the semantic and keyword branches respectively, and $r_{\mathrm{BM25}}$ is the original rank of the passage in the keyword results; together with the damping factor 0.1, this ensures that lower-ranked keyword matches contribute progressively less to the final score. The term $w_{\mathrm{coll}}$ is a boost factor that favors documentation collections most relevant to the type of question being asked, as determined by the query classifier (Section~\ref{sec:workflow}). This allows the system to prioritize the most relevant subset of documentation without excluding any collection entirely. A final filtering step removes low-scoring passages and limits the total context size before passing it to the answer generation step.

\subsection{Agentic Query-Answer Workflow}
\label{sec:workflow}
 
\begin{figure} [ht]
\begin{center}
\begin{tabular}{c} 
\includegraphics[width=\textwidth]{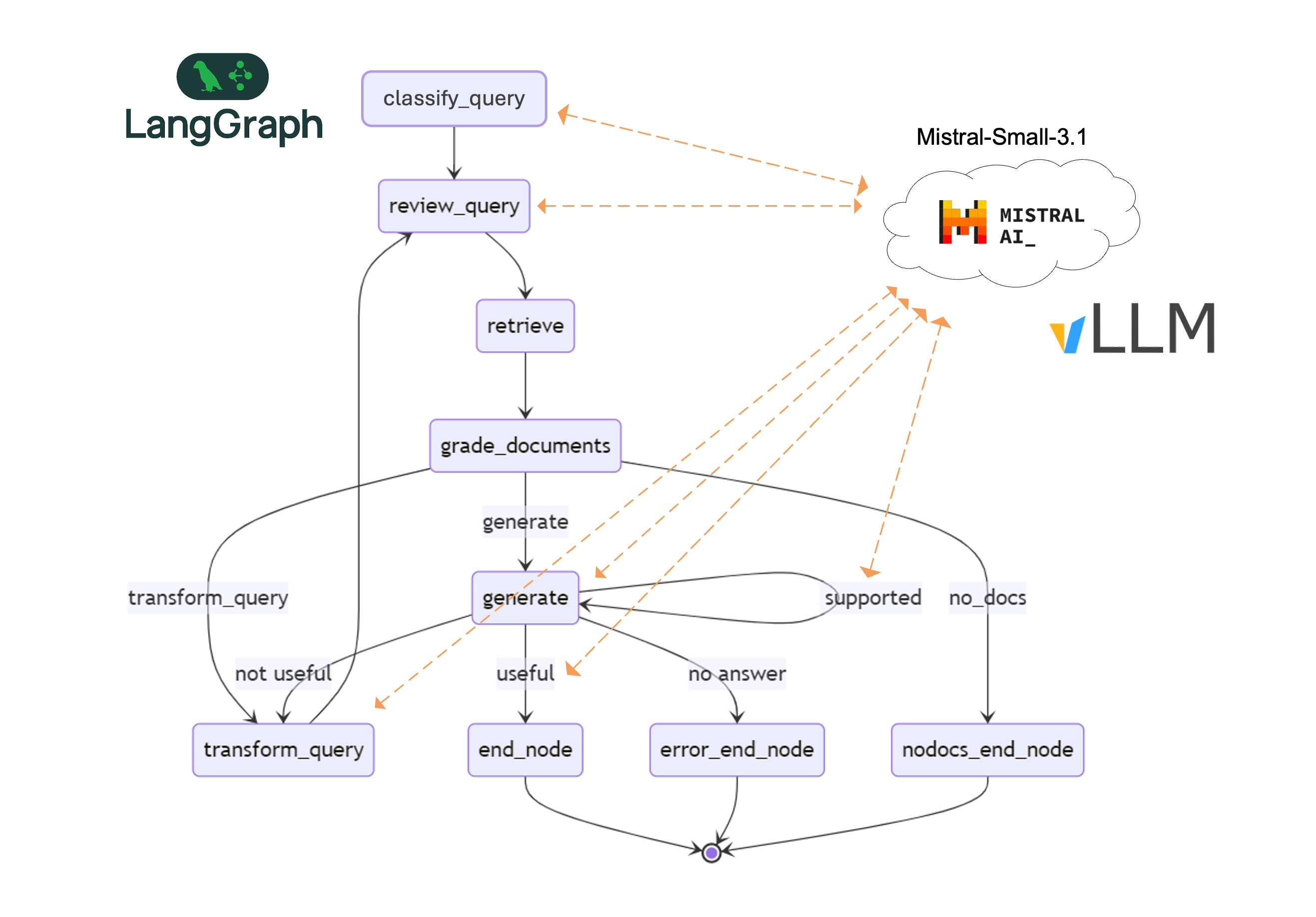}
\end{tabular}
\end{center}
\caption
{ \label{fig:graph} 
ESOFinder agentic query-answer loop summary. The yellow arrows indicate steps that make use of the \texttt{mistral-small-3.1} LLM. The retrieve step follows the strategy presented in Figure~\ref{fig:rag}. }
\end{figure}

The query-to-answer loop is implemented as a self-correcting agentic workflow using LangGraph\footnote{\url{https://github.com/langchain-ai/langgraph}}, a framework for building structured, multi-step LLM pipelines. The workflow performs several LLM-driven steps before and after retrieval, and includes automatic retry logic to improve answer quality. Figure~\ref{fig:graph} summarizes the agentic loop strategy used by ESOFinder.

All LLM calls within the workflow (query classification, question phrasing, message condensation, answer generation, and answer grading) are handled by a single local model, \texttt{Mistral-Small-3.1-24B-Instruct-2503}\footnote{\url{https://mistral.ai/news/mistral-small-3-1}} (hereafter \texttt{mistral-small-3.1}), served via vLLM~\cite{vllm}. Using a single model for all tasks simplifies deployment and ensures consistent behaviour across the pipeline.
 
\textbf{Query understanding:}
Before retrieval begins, two preparatory steps are applied. For multi-turn conversations, the latest user message is condensed into a fully self-contained question that clarifies implicit references to previous turns. The question is then classified into one of seven observatory-specific domains (proposal preparation, instrument configuration, data delivery, pipeline reduction, ALMA, observing support, and general queries) to set the collection weighting ($w_{\mathrm{coll}}$ in Eq.~\ref{eq:fusion}) for retrieval (\texttt{classify\_query}). Next, an alternative phrasing of the question (\texttt{review\_query}) is generated and submitted alongside the original to broaden retrieval coverage.

\textbf{Document retrieval and grading:}
The hybrid retrieval pipeline (Section~\ref{sec:retrieval}) is executed for each question variant, and the resulting candidate documents are merged, deduplicated, and ranked by their final combined score (\texttt{retrieve}). A grading step then discards passages whose combined retrieval score falls below a minimum threshold and retains at most eight documents to be passed to the generation step (\texttt{grade\_documents}). If no documents survive filtering, the query is rewritten (\texttt{transform\_query}) and retrieval is retried up to two additional times; if no relevant documents are found after all retries, a fallback response is returned directly without invoking the generation step (\texttt{nodocs\_end\_node}).

\textbf{Answer generation:}
The \texttt{mistral-small-3.1} LLM generates a structured answer grounded in the retrieved documents, including inline citations and links to the source documentation (\texttt{generate}). All prompts include an ESO-specific terminology glossary to assist the model with observatory acronyms and instrument names.
 
\textbf{Self-correction loop:}
Two grading steps evaluate the generated answer using the \texttt{mistral-small-3.1} LLM: one checks whether the answer is factually grounded in the retrieved documents (\texttt{supported}), and another checks whether it addresses the question asked (\texttt{useful}). If both checks pass, the generated answer is presented to the user (\texttt{end\_node}). If either check fails, the system rewrites the query (\texttt{transform\_query}) and retries the retrieval-generation cycle up to two additional times before returning a fallback response (\texttt{error\_end\_node}). This loop reduces the rate of hallucinated or off-topic answers without requiring human intervention.

\subsection{Deployment and User Interface}
\label{sec:serving}

The ESOFinder pipeline is served through a streaming HTTP backend (FastAPI\footnote{\url{https://fastapi.tiangolo.com}}) that returns intermediate status events and the final answer incrementally, providing users with real-time feedback on system progress. The LLM inference server runs on two GPUs (NVIDIA RTX 6000 Ada Generation) in tensor-parallel mode; the cross-encoder reranker is kept on CPU to prevent memory contention with the LLM inference server.

Users can access ESOFinder in two ways. The primary interface is a web application that supports multi-turn conversations, i.e., the user can ask follow-up questions within the same session, and the system remembers the context of previous exchanges. Multiple independent interfaces can be deployed simultaneously for different user groups, with access currently restricted to ESO staff, fellows, and students during the evaluation phase. In addition, ESOFinder exposes an Application Programming Interface (API) service that is currently integrated with the USD helpdesk system: when a new support ticket is received, the API is queried automatically, and the system's answer is made available to the support astronomer handling the ticket, who may use it to streamline the response process. Both interfaces collect user feedback for system evaluation

\section{ESOFinder's feedback exercise}\label{sec:evaluation}

ESOFinder has undergone three formal feedback exercises since its initial deployment, each informing a subsequent iteration of the system. In all these exercises, users were asked to evaluate each system response using a two-part feedback form: a five-star overall rating, and a qualitative classification selected from a fixed set of options -- \textit{Perfect answer, accurate}; \textit{Correct answer, but missing information}; \textit{Correct answer, but too verbose}; \textit{Correct, but not useful in the context of the question}; \textit{Irrelevant}; \textit{Incorrect answer}; \textit{Confusing}; and \textit{Cannot be answered by ESOFinder}. The star rating provides an overall performance score, while the qualitative labels helps identify the specific nature of each success or failure. The version presented in this paper represents the third generation of the tool, incorporating lessons learned from the first two rounds of evaluation, and is currently undergoing a broader assessment by ESO staff, fellows, and students.

The first feedback exercise was conducted with ESO's user support astronomers, who are responsible for providing support for VLT observations (12 members of the ESO's User Support Group -- USG), using an early version of ESOFinder that employed a simpler agentic workflow, less extensive documentation, and a less refined document retrieval strategy than the architecture described in Section~\ref{sec:architecture}. The overall rating averaged 3.0 out of 5 stars, with negative marks driven primarily by irrelevant or incorrect answers. In approximately one third of the queries, the system's response was judged to be confusing or incorrect. It is worth noting, however, that a fraction of these cases could be traced back to gaps or ambiguities in the underlying documentation rather than failures in the retrieval or generation pipeline. Analysis of the feedback revealed several recurring patterns: questions about specific instruments (such as VISIR) received proportionally more negative reviews than average, the system struggled with convoluted or multi-part questions, and answers were often perceived as overly verbose, which was identified as the primary source of confusion. These findings motivated the incorporation of additional documentation and adjustments to the system prompts to improve answer conciseness, resulting in a second version of the tool with an improved agentic loop and a more diverse knowledge base.

The second feedback exercise was conducted with a broader audience, including staff responsible for maintaining ESO's data pipelines, operating the science archive, and supporting observers throughout the observing process ($\sim 20$ members of the ESO Data Management and Operations -- DMO-- division), yielding an average rating of 3.1 out of 5 stars. The response quality distribution was notably bimodal, with a higher proportion of answers rated either as perfect or as incorrect compared to the first round, suggesting that the system performs well on straightforward queries but still fails on a subset of more challenging ones. As in the first round, a fraction of incorrect answers could be attributed to documentation shortcomings. Specific issues identified included retrieval failures on documents containing large tables, the absence of ALMA documentation from the indexed knowledge base, incomplete coverage of Astronomical Data Query Language (ADQL) query documentation, and Phase~3 documentation being retrieved in response to general queries for which it was not the most appropriate source. The results of this second exercise motivated a substantial redesign of the system, resulting in a third version with an improved agentic loop and a more structured retrieval pipeline, in particular a refined RAG strategy and a reorganisation of the domain-specific document collections.

\begin{figure} [ht]
\begin{center}
\begin{tabular}{cc} 
\includegraphics[width=0.42\textwidth]{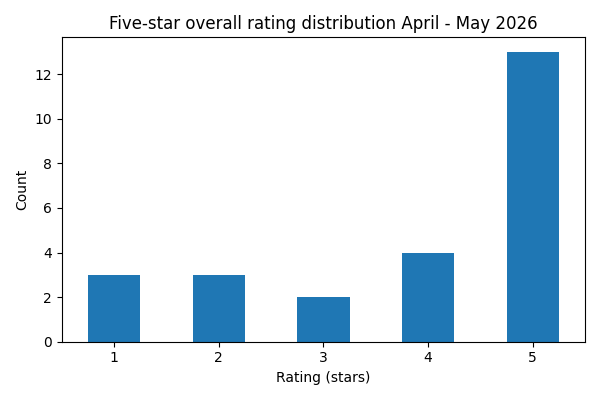} & 
\includegraphics[width=0.54\textwidth]{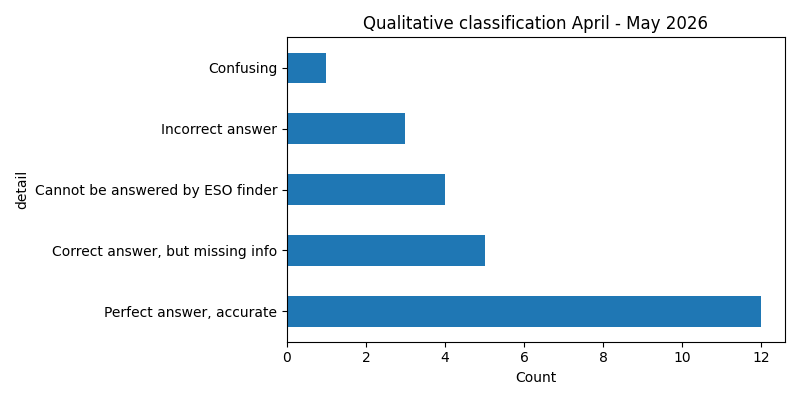}
\end{tabular}
\end{center}
\caption
{ \label{fig:feedback} 
Summary of the third feedback exercise. The left panel shows the distribution of the five-star overall rating, while the right panel shows the more detailed qualitative classification.}
\end{figure}

The current version of ESOFinder, described in this paper, incorporates the improvements driven by the first two feedback exercises and is currently being evaluated by ESO staff astronomers, fellows, and students through the web interface described in Section~\ref{sec:serving}. Preliminary results collected between April and May 2026 indicate a clear improvement in performance: the average star rating has increased to 3.84 out of 5, with a median of 5 stars. When only DMO staff are considered in the feedback evaluation, the average star rating is 4.08, with a median of 5 stars. The majority of detailed feedback responses classify the answers as either \textit{Perfect answer, accurate} or \textit{Correct answer, but missing information}, and the fraction of responses labelled as incorrect has notably reduced compared to previous evaluation rounds. Figure~\ref{fig:feedback} shows the distribution of star ratings and detailed qualitative feedback labels collected during this evaluation period. 

It should be noted that the three evaluation exercises were conducted with different audiences: the first two rounds involved primarily instrument and operations experts from the DMO division, while the current exercise covers a broader and more representative sample of ESO users. This evolution in audience composition is an important caveat when comparing results across rounds, but it also means that the latest evaluation provides a more realistic picture of system performance, as would be expected when used by the broader ESO astronomical community.

under real-world usage conditions.

\section{CONCLUSIONS}
\label{sec:conclusions}

We have presented ESOFinder, a domain-specific question-answering system developed in-house at ESO to assist staff and community astronomers in navigating the observatory's extensive documentation ecosystem. The system is built on a RAG architecture combining hybrid retrieval (semantic search and BM25 keyword indexing), cross-encoder reranking, and a self-correcting agentic workflow powered by a locally hosted open-source LLM. By grounding all responses in a curated, versioned knowledge base, ESOFinder provides concise, reference-linked answers while avoiding the limitations of general-purpose tools, in particular their fixed training cutoff and lack of domain-specific knowledge.

Three successive feedback exercises have guided the iterative development of the system. Each round identified concrete weaknesses in retrieval coverage, documentation completeness, and response quality, and directly motivated targeted improvements to the agentic loop, the RAG strategy, and the organisation of the domain-specific document collections. Preliminary results from the ongoing third evaluation round, covering a broader and more representative user population, show a clear improvement in performance relative to earlier versions, with an average star rating of 3.84 out of 5 and a median of 5 stars, and a shift in the qualitative feedback distribution towards correct answers.

Several limitations remain and will guide future development. Retrieval quality on documents containing large tables and specific information presented as figures (instead of text) requires further work. The ALMA knowledge base, currently maintained as a separate collection, will be expanded and better integrated with the rest of the documentation. Furthermore, the feedback exercise has revealed inconsistencies and gaps in the underlying ESO documentation itself, highlighting the importance of maintaining a well-curated and up-to-date knowledge base.

Looking ahead, we identify several directions for future work. We plan to leverage ESOFinder more systematically as a tool for documentation quality control and improvement, using it to identify contradictions, ambiguities, and outdated content across the documentation on a regular basis. On the technical side, we plan to explore alternative LLM serving strategies and hierarchical multi-agent architectures that could improve both response speed and the system's ability to navigate the layered structure of ESO documentation. We also aim to extend the coverage of the knowledge base by integrating the helpdesk ticket history from the DeskPro system, which would allow ESOFinder to retrieve relevant past tickets and their solutions when support astronomers handle new queries, while ensuring that access to sensitive ticket data remains restricted to authorised USD staff. For this, the use of local LLMs presents a clear advantage, as sensitive information can be fed to the LLM without risking data leaks. In parallel, we are exploring the feasibility of adapting the ESOFinder framework to serve the needs of Paranal operations staff, for example through a dedicated interface tailored to on-site operational documentation and tools used by the telescope and instrument operators. We plan to make the system available to the broader ESO user community, establishing ESOFinder as a standard interface for documentation queries across the observatory.

\acknowledgments 
 
The authors thank all members of the User Support Group (USG) and the Data Management and Operations (DMO) division who participated in the first two feedback exercises, providing detailed and constructive evaluations that directly shaped the development of ESOFinder. We also thank the ESO staff, fellows, and students who are contributing to the ongoing third evaluation round. Their collective effort in testing the system and providing feedback has been invaluable in improving the quality and reliability of the tool.

The authors acknowledge the use of Claude (Anthropic) as a language editing tool during the preparation of this manuscript.

\bibliography{report} 
\bibliographystyle{spiebib} 

\end{document}